\documentclass[12pt,preprint]{aastex}
\usepackage{apjfonts}
\newcommand{\grb}{GRB~130907A}
\begin{document}

\title{ An inverse Compton origin for the  55 GeV photon in the late afterglow of GRB~130907A}
\author{Qing-Wen Tang\altaffilmark{1,3}, Pak-Hin Thomas Tam\altaffilmark{2},
Xiang-Yu Wang\altaffilmark{1,3}}
\affil{$^1$ School of Astronomy and Space Science, Nanjing University, Nanjing, 210093, China \\ xywang@nju.edu.cn \\
$^2$ Institute of Astronomy and Department of Physics, National Tsing Hua University, Hsinchu 30013, Taiwan \\ phtam@phys.nthu.edu.tw \\
$^3$ Key laboratory of Modern Astronomy and Astrophysics (Nanjing University), Ministry of Education, Nanjing 210093, China \\
}
\begin{abstract}
The extended high-energy gamma-ray ($>$100 MeV) emission which
occurs well after the prompt gamma-ray bursts (GRBs) is usually
explained as the afterglow synchrotron radiation. Here we report
the analysis of the Fermi Large Area Telescope observations of
\grb. A 55 GeV photon compatible with the position of the
burst was found at about 5 hours after the prompt phase. The
probability that this photon is associated with \grb~is higher
than 99.96\%. The energy of this photon exceeds the maximum
synchrotron photon energy at this time and its occurrence thus
challenges the synchrotron mechanism as the  origin for the
extended high-energy $>$10~GeV emission. Modeling of the broad-band
spectral energy distribution suggests that such  high energy
photons can be produced by the synchrotron self-Compton emission
of the afterglow.
\end{abstract}


\section{Introduction}
The {\it Fermi}/Large Area Telescope (LAT) observations have
revealed that high-energy gamma-ray emission (i.e., $>$100~MeV) of
$\gamma$-ray bursts (GRBs) often lasts much longer than the prompt
KeV/MeV burst \citep{2013ApJS..209...11A}. Such behavior has also
been seen before the launch of {\it Fermi}, {e.g.,} an $\sim$18
GeV photon was observed by $CGRO/$EGRET 90 minutes after the
prompt burst phase~\citep{1994Natur.372..652H}. The usually
discussed scenario for this temporally extended high-energy
emission is the afterglow synchrotron model, where electrons are
accelerated by forward shocks expanding into the circumburst
medium and produce $>100$ MeV photons via synchrotron radiation
~\citep{2009MNRAS.400L..75K,2010MNRAS.409..226K,2010MNRAS.403..926G,2010ApJ...712.1232W,He2011}.
This model works well for the high-energy emission above 100~MeV up to about 10~GeV.
However, since the synchrotron radiation has a maximum photon
energy (typically $\sim50$ MeV in the rest frame of the shock) and
the forward shock Lorentz factor decreases with time, it is
difficult to explain $>$10 GeV photons detected during the
afterglow
phase~\citep{2010ApJ...718L..63P,2011MNRAS.412..522B,2012ApJ...749...80S,2013ApJ...771L..33W}.

In the past five years, several $>$10 GeV photons have been
observed by the $Fermi$/LAT well after the prompt burst.
Interestingly, the time-resolved spectra of the LAT emission of
GRB 130427A shows a signature of spectral hardening at high
energies above several GeV \citep{2013ApJ...771L..13T}, which is
consistent with the synchrotron self-Compton (SSC) emission of the
afterglow~\citep{2013ApJ...773L..20L,2013ApJ...776...95F}, as has
been predicted for a long while
~\citep{1994MNRAS.269L..41M,2001ApJ...559..110Z,2001ApJ...548..787S,2009MNRAS.396.1163Z,2009ApJ...703...60X}.
The authors of a recent study~\citep{2014Sci...343...42A} do not
claim such an extra hard spectral component at highest energies,
but signature for a  hard  spectrum especially at late times can
be seen in Fig. S1 in the supplementary materials of their paper.
Signature of flattening in the light curves of LAT emission was
also seen in some GRBs, such as GRB 090926A
\citep{2013ApJS..209...11A}, and was interpreted as the SSC
emission by~\citet{2013ApJ...771L..33W}.

In this work, we made use of the publicly available LAT data to
analyze the high  energy emission from \grb.  We describe the
properties of \grb~in section 2. The LAT analysis results are given
in section 3, and  the interpretation  of the high-energy emission
is given in section 4.

\section{Properties of \grb}

\grb~triggered {\it Swift}/BAT at 21:41:13.09 UT on
Sep~7,~2013~
\citep[hereafter $T_\mathrm{0}$;][]{2013GCN..15183...1P}. Its bright prompt emission was also
detected by other satellites, including
Konus-Wind~\citep{2013GCN..15203...1G} and
INTEGRAL~\citep{2013GCN..15204...1S}. Several ground-based
observations followed, which allowed a rapid determination of the
burst location and redshift~\citep[i.e.,
$z=1.238$;][]{2013GCN..15187...1D}, as well as extensive broad band
afterglow monitoring from radio to $\gamma$-rays, e.g.,
VLA~\citep{2013GCN..15200...1C} and
Skynet~\citep{2013GCN..15191...1T,2013GCN..15193...1T}. The
measurement by Konus-Wind gives a fluence of
$(7.9\pm0.5)\times10^{-4}$erg cm$^2$ from 20 keV to 10 MeV. At
redshift of $z=1.238$, the isotropic energy of the burst is $E_{\rm
iso}=(3.3\pm0.1)\times10^{54}$~erg. The burst duration is about
250 seconds as seen in the 15$-$150keV band~\citep{2013GCN..15204...1S}.

{\it Swift}/X-ray Telescope (XRT) began observing the GRB at
$T_\mathrm{0}+$ 66.6 s and found a bright, uncatalogued X-ray source
located at R.A.$=+14^h 23^m 34.22^s$, dec$=+45^{\circ} 36^{'}
33.8^{"}$(J2000), with a 90\% containment error circle of radius of $5^{"}$~\citep{2013GCN..15183...1P}. This GRB position was used in the
analysis presented in the following.

\section{Data analysis and results}
\subsection{LAT data analysis}
Both {\it Fermi} GBM and LAT did not trigger on \grb~due to the South
Atlantic Anomaly (SAA) passage.  Fermi resumed data taking at
$\sim T_0+$1.3~ks after it left the SAA, when the burst
was $\sim$150$^{\circ}$ from the LAT boresight. It only entered
the LAT field-of-view (FoV) at $\sim T_0+$3.4~ks.
Preliminary analysis showed that \grb~was detected by the LAT and
a $\sim$55~GeV photon compatible with the GRB position was
observed at $\sim T_0+$18~ks~\citep{2013GCN..15196...1V}.

We analyzed the LAT data from $T_0$+3~ks to $T_0$+80~ks, which are
available at the  Fermi Science Support Center. The Fermi Science
Tools v9r31p1 package was used to analyze the data between 100~MeV
and 60~GeV. As recommended by the LAT team for this time scale,
events of the P7SOURCE$\_$V6 class were used. We selected all the
events within a region of interest (ROI) with a radius of
10$^{\circ}$ around the position of the GRB, excluding times when
any part of the ROI was at a zenith angle $>100^{\circ}$. We
constructed a background source model including the nearby 33
point sources in the second Fermi
catalog~\citep{2012yCat..21990031N}, as well as the galactic
diffuse emission (gal$\_$2yearp7v6$\_$v0.fits) and the isotropic
component (iso$\_$p7v6source.txt). A simple power law
spectrum is assumed for the LAT emission from \grb, which is
described by $N(E)=N_0(E/E_0)^{-\Gamma}$, where $\Gamma$ is the
photon index.

First, an unbinned maximum-likelihood analysis was performed for
the period $T_\mathrm{0}+$3~ks to $T_\mathrm{0}+$20~ks. We
obtained a test-statistic (TS; the square root of the TS is
approximately equal to the detection significance for a given
source, see Mattox et al., 1996) value of 41.3, corresponding to a detection significance of 6.4, consistent with the value given in Vianello et al. (2013). This analysis also gave a photon index of $-$1.9$\pm$0.2 and an average photon flux of
(5.5$\pm$1.8)$\times$10$^{-7}$~cm$^{-2}$s$^{-1}$. We note the
detection of a 54.4~GeV photon at 17218s after $T_0$, and we identify this photon as the one first mentioned in Vianello et al. (2013). We also
searched through the period $T_\mathrm{0}+$20~ks to
$T_\mathrm{0}+$80~ks and did not detect any significant emission.
We put a 90\% confidence level (c.l.) upper limit of the photon flux at 1.2$\times$10$^{-7}$ (or
6.1$\times$10$^{-8}$)~cm$^{-2}$s$^{-1}$ in this time interval,
assuming $\Gamma=$2.0 (or 1.5).

To firmly establish the association of the 54.4~GeV photon with \grb, we
estimated the probability using two methods. First we used the
{\it gtsrcprob} tool, which used the likelihood analysis as described above to assign to each photon in the ROI a
probability that such photon is associated  with \grb~(instead of
coming from the background). There are 6 photons
with probability of $>$90\% and 11 photons with
probability of $>$50\% from $T_\mathrm{0}+$3~ks to
$T_\mathrm{0}+$20~ks. In particular, the 54.4 GeV photon at 17218
seconds has a probability of $>$99.998\% to be associated with
\grb. Secondly, we counted the number of 50--500 GeV photons from 5
years of LAT observations from the direction of GRB 130907A (using
the point-spread-function of 0$\fdg$8 for a 50 GeV photon), which
is three. Therefore, the probability to obtain one 50--500 GeV
background photon from this direction over a time interval of
20~ks is about $3.8\times10^{-4}$. In other words, the probability
that  this photon is associated with \grb~is higher than
{99.96\%}.

The energy resolution of 10$-$100 GeV photons is on the order of
10\%, therefore we will also  designate this highest energy photon
as a 55~GeV photon. At $z=1.238$, the intrinsic photon energy of
the 55~GeV photon would be 123~GeV, putting it as one of the few
very high-energy $\gamma$-ray photon originated from a GRB, after
those from GRB~080916C~\citep{pass8_grbs} and
GRB~130427A~\citep{2013ApJ...771L..13T}.

From $T_\mathrm{0}+$3~ks to $T_\mathrm{0}+$20~ks, the GRB was in
the LAT FoV only during two intervals: 3000--6000s and
14000--20000s after $T_\mathrm{0}$. During each interval, \grb~was detected significantly,
 i.e., a TS value of $>$20 was obtained. We note that the detecton in the second interval is dominated by the single 55~GeV photon.
For each of the above time interval, three or four energy bins were defined when constructing the spectral energy distribution (SED) in the LAT band.
For those fits without well-constrained photon index,
we fixed it to be $\Gamma=2.0$ (the value obtained in
the full-energy fit for the period from 3~ks to 80~ks), as well as $\Gamma=1.5$.
For those energy bins in which the TS value is smaller than 9, 90\% c.l. upper limits are given, again assuming $\Gamma=2.0$ (or $\Gamma=1.5$).
The results are summarized in Table~\ref{flux-analysis}.

The LAT light
curve can be fitted with a single power law with a slope of
$-$1.13$\pm$0.57 as shown in Fig.~\ref{lightcurve}. Photon flux values are given here because energy flux depends largely on the photon indices that are sometimes not well constrained.

\subsection{XRT data analysis}
We extracted {\it Swift}/XRT lightcurve and spectra
during the LAT observations using the standard HEASOFT reduction pipelines
and the {\it Swift}/XRT repository~\citep{evan07,2009MNRAS.397.1177E}. The XRT observations of \grb~started in window timing (WT) mode and data were taken in
the photon counting (PC) mode combined
with WT mode since $\sim T_\mathrm{0}+$7000s. The PC spectrum can be fitted by an absorbed power-law
model with a photon index 1.67$\pm$0.19 in the time interval
from 3000 to 10000 seconds since $T_\mathrm{0}$, and 1.79$\pm$0.12
from 14000 to 20000 seconds since $T_\mathrm{0}$.
The X-ray light curve can be fitted with a
smoothed broken power law with a break at $\sim$19.6ks, the pre-break decay index is $-1.05\pm0.07$, and the post-break decay index is $-2.43\pm0.07$ as shown in Fig.~\ref{lightcurve}.

\section{Interpretations and Discussion}
In this section, we will investigate the origin of the highest energy
photon from \grb~and assist the study by modeling the
multi-band (from radio to GeV bands) data of this burst.
Hereafter, we denote by $Q_x$ the value of the quantity $Q$ in
units of $10^x$.
\subsection{What is the origin of the highest energy photon above 10~GeV? }
By equating the synchrotron cooling time in the magnetic field
with the Lamour time of the electrons, one can obtain the Lorentz
factors of the maximal energy electrons, which is
$\gamma_{e,max}\propto B^{-1/2}$, where $B$ is the strength of the
magnetic field. In the same magnetic field, the maximal
synchrotron photon energy produced by these electrons is about 50
MeV in the shock co-moving frame. Considering the bulk motion
boosting by the forward shock which has a Lorentz factor
$\Gamma_{ext}(t)$, the maximal synchrotron photon energy is $\sim 50
\Gamma_{ext}(t)${\rm MeV}. As the Lorentz factor of the forward shock
decreases significantly at late times, one would expect
$E_{max,syn}\la 1.3(E_{54}/n_0)^{1/8}T_4^{-3/8}${\rm GeV}. The
presence of a 54.4{\rm GeV} photon at $\sim T_\mathrm{0}+$5 hours is incompatible with
a synchrotron origin even for the assumption about the fastest
acceleration.
It is more likely to originate from the inverse-Compton process,
as also suggested by the detailed modeling of $>$10~GeV photons
from GRB~130427A and other
bursts (e.g.~\citet{2013ApJ...771L..33W,2013ApJ...773L..20L}). The
modeling of the broad-band SED, as shown
in the following section, supports this explanation. Such a
high-energy SSC emission has been
predicted to be present in the high-energy afterglow emission for
over a decade, e.g.~\citet{2001ApJ...559..110Z,2001ApJ...548..787S}.
\subsection{The model}
In the standard synchrotron afterglow spectrum,  there are
three break frequencies, $\nu_a$, $\nu_m$, and $\nu_c$, which are
caused by synchrotron self-absorption, electron injection, and
electron cooling, respectively. According to, e.g.
~\citet{1998ApJ...497L..17S} and ~\citet{1999ApJ...523..177W}, the
cooling Lorentz factor and the minimum Lorentz factor of electrons
in forward shocks are given by
\begin{equation}
\gamma_m=7.5\times 10^{2} f(p) \epsilon_{e,-1} E_{54}^{1/8}
n_0^{-1/8}(1+z)^{3/8} T_4^{-3/8},
\end{equation}
and
\begin{equation}
\gamma_c=3.5\times 10^6 E_{54}^{-3/8} n_0^{-5/8}
\epsilon_{B,-5}^{-1}(1+z)^{-1/8}(1+Y_c)^{-1} T_4^{1/8},
\end{equation}
where $E$ is the kinetic energy of the spherical shock,
$\epsilon_e$ is the fraction of the shock energy that goes into
the electrons, $\epsilon_B$ is fraction of the shock energy that
goes into  the magnetic fields, $n$ is the particle number density
of the uniform medium, and $Y_c$ is the Compton parameter for
electrons of energy $\gamma_c$ and $f_p = 6(p-2)/(p-1)$), with $p$
being the electron index. These three characteristic frequencies
are given by
\begin{equation}
\nu_a=8.9\times 10^{9} f^{-1}(p) \epsilon_{e,-1}^{-1} E_{54}^{1/5}
n_0^{3/5}\epsilon_{B,-5}^{1/5}(1+z)^{-1} Hz,
\end{equation}
\begin{equation}
\nu_m=1.2\times 10^{12} f^2(p) \epsilon_{e,-1}^2 E_{54}^{1/2}
\epsilon_{B,-5}^{1/2}(1+z)^{1/2} T_4^{-3/2} Hz,
\end{equation}
and
\begin{equation}
\nu_c=2.5\times 10^{19} E_{54}^{-1/2}
n_0^{-1}\epsilon_{B,-5}^{-3/2} (1+Y_c)^{-2} (1+z)^{-1/2} T_4^{-1/2}
Hz.
\end{equation}
We can also get the peak flux of the synchrotron emission by
\begin{equation}
F_{\nu,m}=3.5\times 10^4 E_{54} n_0^{1/2}\epsilon_{B,-5}^{1/2}
D_{28}^{-2}(1+z) \mu {\rm Jy},
\end{equation}
where $D$ is the luminosity distance. For the SSC emission, we get the corresponding quantities
according to~\citet{2001ApJ...548..787S}, i.e.
\begin{equation}
h\nu_m^{IC}=5.4 f^4(p) \epsilon_{e,-1}^4 E_{54}^{3/4}
\epsilon_{B,-5}^{1/2}(1+z)^{5/4} T_4^{-9/4} {\rm keV},
\end{equation}
\begin{equation}
h\nu_c^{IC}=2.5\times 10^{6} E_{54}^{-5/4}
n_0^{-9/4}\epsilon_{B,-5}^{-7/2} (1+Y_c)^{-4} (1+z)^{-3/4} T_4^{-1/4}
{\rm TeV},
\end{equation}
and
\begin{equation}
F_{\nu,m}^{IC}=5.6\times 10^{-3} E_{54}^{5/4}
n_0^{5/4}\epsilon_{B,-5}^{1/2} D_{28}^{-2}(1+z)^{3/4}T_4^{1/4} \mu
{\rm Jy}.
\end{equation}
{Due to the fact that the inverse-Compton scatterings between
$\gamma_c$ electrons and the peak energy photons at frequency
$\nu_c$ are typically in the Klein-Nishina (KN) regime, the SSC
emission indeed peaks at
\begin{equation}
h \nu^{IC}_{peak}=\Gamma_{ext} m_e c^2
\gamma_c=44(1+Y_c)^{-1}E_{54}^{-1/4}n_0^{-3/4}\epsilon_{B,-5}^{-1}(1+z)^{1/4}T_4^{-1/4}{\rm
TeV},
\end{equation}
where $\Gamma_{ext}$ is the bulk Lorentz factor of the external
forward shock. Above the peak $h \nu^{IC}_{peak}$, the spectrum
has the form similar to   Eq.50 in Nakar et al. (2009).} In order
to calculate $Y_c$, we need to consider the SSC emission in the KN
scattering regime. Following Wang et al. (2010), we define a
critical frequency, above which the scatterings with electrons of
energy $\gamma_c$ just enter the KN scattering regime, i.e.
 \begin{equation}
\nu_{{\rm KN}}(\gamma_c)=8.7\times 10^{14} E_{54}^{1/2} n_0^{-1/2}\epsilon_{B,-5}^{-1/2} (1+Y_c) (1+z)^{-1/2} T_4^{-1/2} Hz
\end{equation}
In a broad parameter space, we find $\nu_m<\nu_{{\rm
KN}}(\gamma_c)<\nu_c$, thus following~\citet{2010ApJ...712.1232W},
we have
$Y_c(1+Y_c)=\frac{\epsilon_e}{\epsilon_B}(\frac{\gamma_c}{\gamma_m})^{2-p}(\frac{\nu_{{\rm
KN}}(\gamma_c)}{\nu_c})^{(3-p)/2}$ ({which is also consistent with
eq. (47) in Nakar et al. 2009)}. {Therefore, $Y_c$ is in the range
from 4 to 6 in the time range between 1ks and 20ks after the burst
for the reference parameter values of
$\epsilon_{e,-1}=\epsilon_{B,-5}=E_{54}=n_0=1$.}

\subsection{Modeling the multi-wavelength emission}
The multi-wavelength light curves are shown in Fig.~1. The X-ray
light curves (3-10keV) initially decays as $t^{-1.05\pm0.07}$ in
the first 20ks and then steepens to a faster decay as
$t^{-2.43\pm0.07}$. The early X-ray decay slope $-1.05\pm0.07$ and
the spectral index $\beta_{\rm X}=0.67\pm0.19$ are consistent with
$\alpha=-(3p-3)/4$ and $\beta=-(p-1)/2$ ($F_ \nu\propto
t^{\alpha}\nu^{\beta}$) predicted by the standard afterglow
synchrotron emission in the slow-cooling regime (i.e.
$\nu_m<\nu_{\rm X}<\nu_c$) with $p\simeq2.3$, assuming that the
blast wave expands into a constant density circumburst medium.
{The wind medium scenario is not favored because the predicted
decay slope in the wind medium scenario $\alpha_{w}=-(3p-1)/4$ is
too steep}. By fitting the {\it I}-band data from
Skynet~\citep{2013GCN..15193...1T} and {\it i}-band data from
RATIR~\citep[][converted to I-band data according to Blanton \&
Roweis,
2007]{2013GCN..15208...1B,2013GCN..15209...1B,2013GCN..15192...1L},
we got the decay slope $-0.87\pm0.09$, which is consistent with
$\sim$0.9 in~\citet{2013GCN..15193...1T}. The decay slope is
consistent with the predicted slope $\alpha=-(3p-3)/4$ when the
optical frequency $\nu_{\rm O}$ also locates at $\nu_m<\nu_{\rm
O}<\nu_c$. It is puzzling that the optical light curve does not
show a break as X-rays. We speculate that there might be some
extra contribution to the late-time optical flux, such as the host
galaxy or another possible outflow component. Such chromatic break
behavior has been seen before in other GRBs as well, e.g.
\citet{2006MNRAS.369.2059P}.

The LAT emission decays as $t^{-1.13\pm0.57}$ with a large error
bar in the decay slope due to small number of high-energy photons.
The decay slope is consistent with the predicted slope
$\alpha=-(3p-2)/4$, as the LAT energy window is above the cooling
frequency $\nu_c$. Note that the time evolution of the Compton
parameter (defined as $Y(\gamma_*)$ in
~\citet{2010ApJ...712.1232W}) for the high-energy emission could
steepen the LAT synchrotron decaying light curve, since the
high-energy synchrotron flux scales as $1/(1+Y(\gamma_*))$. {The
value of $Y(\gamma_*)$ is $\simeq 2$ at a few ks.} However,
because the measured decay slope has a large error (i.e. $-1.13\pm
0.57$), such effect is hard to be discerned from the data.

Below we model the
multi-band SED with the synchrotron plus SSC emission model.

The LAT flux at $h\nu=100$ {\rm MeV} is about $(6.0\pm3.3)\times
10^{-4}\mu {\rm Jy}$ around 6000s after the burst. Since at
$h\nu=100 {\rm MeV}$ the flux is dominated by the synchrotron
emission, we have
\begin{equation}
F_\nu({\rm 100{\rm MeV}}, t=6{\rm
ks})=F_{\nu,m}(\frac{\nu_c}{\nu_m})^{-(p-1)/2}(\frac{100{\rm MeV}}{h\nu_c})^{-p/2}
=(6.0\pm3.3)\times 10^{-4}\mu {\rm Jy}.
\end{equation}
The X-ray  flux density at 1 keV around 17ks is
about $(11.6\pm2.1) \mu {\rm Jy}$ which was derived from the Swift XRT
data, as discussed above, and we have
\begin{equation}
F_\nu(1{\rm keV}, t=17{\rm
ks})=F_{\nu,m}(\frac{1{\rm keV}}{h\nu_m})^{-(p-1)/2}=(11.6\pm2.1) \mu
{\rm Jy}.
\end{equation}
VLA observations started at $\sim T_\mathrm{0}+$14ks and ended at $\sim T_\mathrm{0}+$20ks at $\nu=24.5$GHz,
the flux is ($1.2\pm0.09$)m{\rm Jy}~\citep{2013GCN..15200...1C}. Since the high frequency of radio band locates
between the self-absorption frequency $\nu_a$ and $\nu_m$, we have
\begin{equation}
F_\nu(24.5{\rm GHz}, t=4{\rm
hr})=F_{\nu,m}(\frac{24.5{\rm GHz}}{\nu_m})^{1/3}\simeq(1.2\pm0.09) m{\rm Jy} .
\end{equation}
The flux at  10{\rm GeV} at 17000s after the burst should be dominated
by the SSC emission as discussed above. With a flux of (1.75$\pm$1.48) $\times
10^{-5}\mu {\rm Jy}$ at 10 GeV, we have
($h\nu_m^{IC}$<10{\rm GeV}$<h\nu_c^{IC}$)
\begin{equation}
F_\nu(10{\rm GeV}, t=17{\rm
ks})=F_{\nu,m}^{IC}(\frac{10{\rm GeV}}{h\nu_m^{IC}})^{-(p-1)/2}=(1.75\pm1.48)
\times 10^{-5}\mu {\rm Jy} .
\end{equation}
We find that the following parameter values are consistent with
the observed data: $E_{54}=0.75$, $\epsilon_{e,-1}=3.2$,
$\epsilon_{B,-5}=0.55$, $n={\rm 1.8 cm^{-3}}$. {The obtained value
of $\epsilon_B$ is quite small and the obtained value of the
kinetic energy implies a very large efficiency in producing
gamma-rays}. {For these parameter values, we find that $Y_c$ is
about 12-16 in the time between 1ks and 20ks after the burst, and
the assumption ${\nu_m}<{\nu_{KN}(\gamma_c)}<{\nu_c}$ holds in
this time range. Since in our case the forward shock electrons are
in the slow-cooling regime (i.e. $\gamma_m<\gamma_c$), $Y_c$
roughly reflects the total luminosity ratio between the SSC
component and the synchrotron
component~\citep{2009ApJ...703..675N}. The fluence emitted by the
SSC component can be estimated from the spectral energy
distribution in Fig.2, which is roughly $F\sim2\times10^{-5} {\rm
erg \,cm^{-2}}$, so the  energy radiated in the SSC component is
about $8\times10^{52} {\rm erg}$. The fluence in the synchrotron
component is about a factor of 10 smaller than the SSC component,
consistent with the estimated value of $Y_c$. Thus, the total
energy in the afterglow emission is about one order of magnitude
lower than the inferred kinetic energy of the forward shock, which
is $E=7.5\times10^{53} {\rm erg}$.

Guided by the above parameter values, we model the multi-band SED
with the synchrotron plus SSC emission model, as shown in
Fig.~\ref{spectrum}. For the synchrotron component, we use a
series of  smoothly connected  power-laws with two breaks at
$\nu_m$ and $\nu_c$ as well as an exponential cutoff at
$E_\mathrm{max,syn}$. {And for the SSC component, we use a series
of smoothly connected power-laws with breaks at $\nu_m^{IC}$ and
$\nu_{peak}^{IC}$}. The results favor that the $>10$~GeV emission
is produced by the SSC component.

\subsection{Discussions}
{{\it Fermi}/LAT has detected long-lasting high-energy photons
(>100 MeV) from about 60 gamma-ray bursts (GRBs) as of Feb
2014~\footnote{\url{http://www.asdc.asi.it/grblat/}}, with the
highest energy photons reaching about 100 GeV.}  One widely
discussed scenario is that this emission is the afterglow
synchrotron emission produced by electrons accelerated in the
forward shocks. It has been pointed out that, although this
scenario can explain the extended 100MeV-GeV emission very well,
it has difficulty in explaining $>10$ GeV photons, especially in
the late afterglow, since the maximum synchrotron photon energy is
limited
~\citep{2010ApJ...718L..63P,2011MNRAS.412..522B,2012ApJ...749...80S,2013ApJ...771L..33W}.
 The  late $>10$ GeV emission could in principle be
produced by the afterglow SSC emission, as indeed has been invoked
to explain the late $>10$ GeV emission in e.g. GRB130427A ~\citep{2013ApJ...771L..13T,2013ApJ...776...95F,2013ApJ...773L..20L,2013MNRAS.436.3106P} and GRB090926A~\citep{2013ApJ...771L..33W}.

In this paper, we report the analysis of the Fermi/LAT
observations of \grb~ and the detection of one 55 GeV photon
compatible with the position of \grb~ at about 5 hours after
the burst. The probability that this photon is associated with
\grb~ was found to be higher than 99.96\%. At this late time, the
energy of this photon exceeds the maximum synchrotron photon
energy. We modeled the broad-band SED of
the afterglow emission and find that this high energy photon is
consistent with the SSC emission of the afterglow.

\acknowledgments We thank the anonymous referee who has helped to improve the manuscript. We also thank Albert Kong and Ruo-Yu Liu for useful discussion. This work made use of data supplied by the Fermi Science Support Center and the UK Swift Science Data Centre at the University of Leicester. This work is supported by the 973 program under
grant 2014CB845800, the NSFC under grants 11273016 and 11033002,
and the Excellent Youth Foundation of Jiangsu Province (BK2012011).
PHT is supported by the Ministry of Science and Technology of the Republic
of China (Taiwan) through grant 101-2112-M-007-022-MY3.

\begin{table}
\centering
\caption{LAT analysis results of \grb. \label{flux-analysis}}
\begin{tabular}{cccccccc}
    \hline\hline
    Time since $T_0$ & Energy & TS value & $\Gamma$\tablenotemark{a} & Flux$\tablenotemark{b}$   \\
    (s)                         & (GeV)  &          &          & (10$^{-7}$~ph~cm$^2$~s$^{-1}$)  \\
    \hline
3000--4000  &   0.1--60       &   23.1   &   2.4$\pm$0.5              & 29.0$\pm$14.9   \\
\hline
4000--6000  &   0.1--60       &   16.4   &   2.3$\pm$0.6              &  9.2$\pm$5.4   \\
\hline
14000--20000&   0.1--60       &   24.3   &   2.0(1.5)  & 4.5$\pm$3.3(2.3$\pm$2.0)   \\
\hline
3000--6000  &   0.1--0.2      &  9.5     &   3.4$\pm$1.7     & 6.9$\pm$4.0   \\
...         &   0.2--0.5      &  10.0    &   2.0(1.5)             & 3.4$\pm$1.0(3.3$\pm$0.8)   \\
...         &   0.5--1.1      &  10.0    &   2.0(1.5)             & 1.7$\pm$1.1(1.6$\pm$1.1)    \\
...         &   1.1--60      &  0.0        &   2.0(1.5)             &  $<$1.0($<$1.0)           \\
\hline
14000--20000&   0.1--0.5      &  0.8     &   2.0(1.5)             &  $<$11.8($<$11.3)       \\
...         &   0.5--1.1     &  0.0        &   2.0(1.5)             &  $<$0.9($<$0.9)         \\
...         &   1.1--60      &  27.7     &   2.0(1.5)             &  0.6$\pm$0.5(0.6$\pm$0.5)   \\
    \hline
\end{tabular}
\tablenotetext{a}{~Power-law Index. For those fits in which the index was not well-constrained, a fixed value of 2.0(or 1.5) was assumed.}
\tablenotetext{b}{~Values preceeded by the "$<$" sign indicates 90\% c.l. upper limits. Values in parentheses are derived flux by assuming $\Gamma=1.5$}
\end{table}

      \begin{figure*}
    \epsscale{.6}
   \plotone{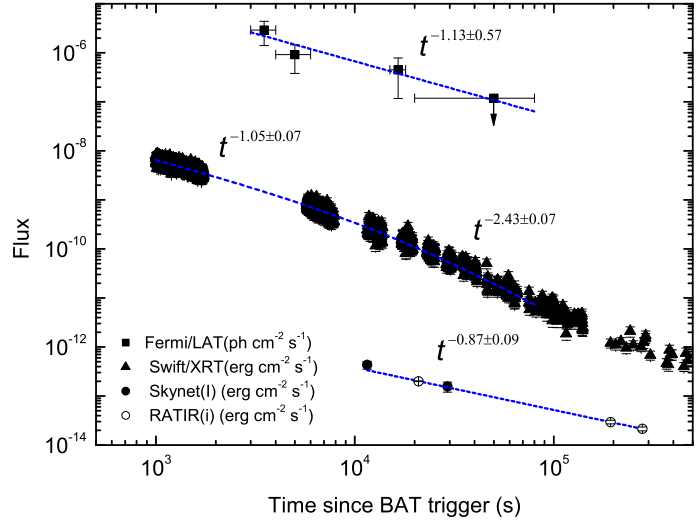}
      \caption{Multi-wavelength lightcurves of \grb. The best fit power-law (for $\gamma$-rays and optical band)
      or smoothly broken power-law (for X-rays) functions for the corresponding times (indicated by the start and end times of the lines) are shown. Flux from RATIR $i$-band observations were converted to the corresponding $I$-band flux following Blanton \& Roweis (2007), to match the Skynet $I$-band observations.}
         \label{lightcurve}
   \end{figure*}

   \begin{figure*}
    \epsscale{.6}
    \plotone{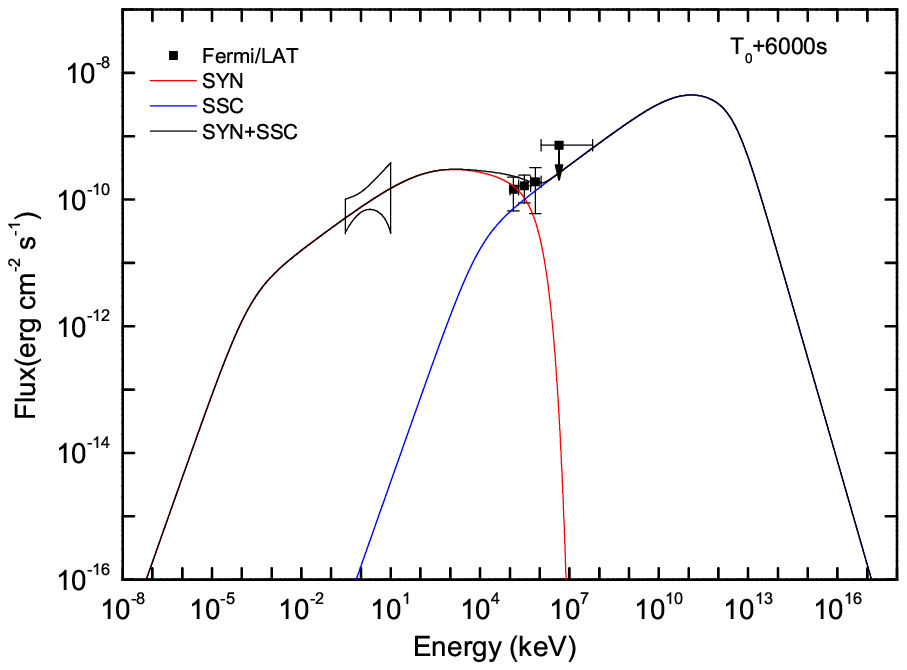}
    \plotone{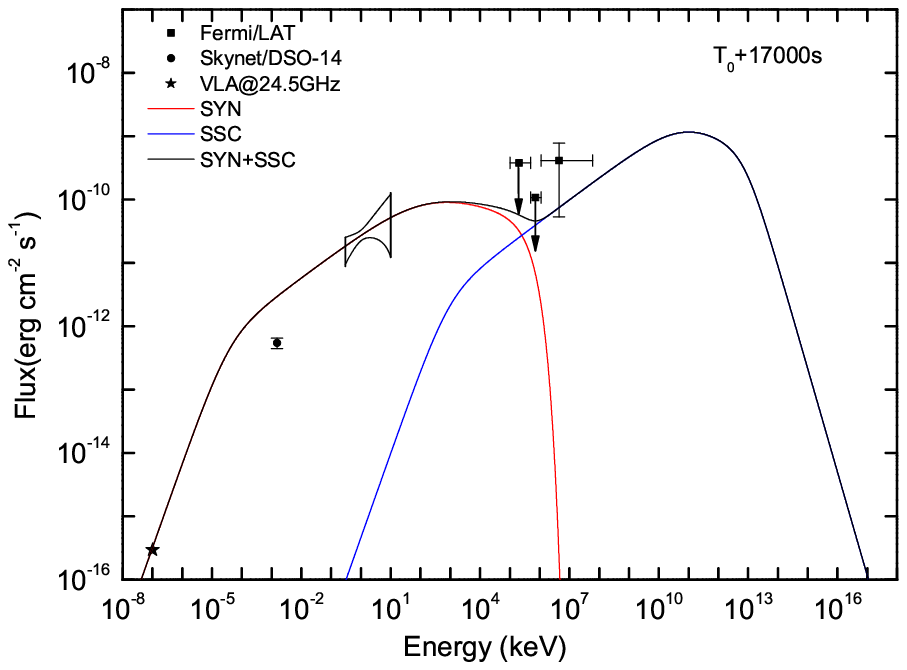}
      \caption{The broadband SEDs at around 6000s (upper panel) and 17000s (lower panel) after the burst.
      The LAT data points were derived from observations during 3ks--6ks and 14ks--20ks after $T_0$, respectively,
      as presented in Table~1. Upper limits assuming $\Gamma=2.0$ are shown. The butterflies represent the best-fit
      spectra in the 0.3-10 keV range derived from the absorbed power-law model in two time intervals after $T_0$:
      3ks--10ks and 14ks--20ks. Optical and radio flux are taken
      from Skynet~\citep{2013GCN..15191...1T,2013GCN..15193...1T} and VLA~\citep{2013GCN..15200...1C} observations,
      respectively. The red and blue dotted lines represent the model synchrotron component and
      SSC component respectively and the black solid lines represent the sum of them. Discrepancy between
      the optical ($I$ band) data and the model flux in the bottom panel can be accounted for by the extinction of the host galaxy
      using the column density given in the XRT fits. The model lines are
      calculated based on the following parameter values: $E=0.5\times 10^{54}$ ergs, $\epsilon_e=0.35$,
      $\epsilon_B=0.44\times 10^{-5}$, $n=1.78 cm^{-3}$. {Note that the parameter values here are slightly different
      from those in the text because smoothed broken power-law functions, instead of power-law functions with sharp breaks,
      are used for the fit  of the synchrotron and SSC spectra of the SED shown here.}}
         \label{spectrum}
   \end{figure*}

\end{document}